\documentclass[%
 reprint,
 superscriptaddress,
 amsmath,amssymb,
 aps,
]{revtex4-2}
\usepackage{graphicx}
\usepackage{dcolumn}
\usepackage{bm}
\usepackage{subcaption}

\usepackage[utf8]{inputenc}
\usepackage{mwe}

\usepackage{xcolor}
\usepackage{mathtools}

\usepackage{physics}
\usepackage[normalem]{ulem}
\usepackage{lineno}
\usepackage{soul} 
\usepackage{float} 

\usepackage{hyperref} 
\hypersetup{colorlinks=true, citecolor=blue}

\newcommand{\etal}[1][\space]{\textit{et al.}#1} 
\newcommand{\ute}[1][\space]{UTe$_2$#1} 
\newcommand{\EE}[1]{\times 10^{#1}} 
\newcommand{\ohm}[1][\space]{$\Omega$#1} 

\begin{document}

\preprint{APS/123-QED}

\title{Evidence of anisotropic surface state residual loss in \ute amid isotropic low-energy excitations}

\author{Arthur Carlton-Jones}
\email[Corresponding author: ]{acarlton@terpmail.umd.edu}
\affiliation{Maryland Quantum Materials Center, Physics Department,\\
University of Maryland, College Park, MD 20742-4111  USA}
\author{Jared Z. Dans}
\affiliation{Maryland Quantum Materials Center, Physics Department,\\
University of Maryland, College Park, MD 20742-4111  USA}
\author{Gicela Saucedo Salas}
\affiliation{Maryland Quantum Materials Center, Physics Department,\\
University of Maryland, College Park, MD 20742-4111  USA}
\author{Shanta R. Saha}
\affiliation{Maryland Quantum Materials Center, Physics Department,\\
University of Maryland, College Park, MD 20742-4111  USA}
\author{Johnpierre Paglione}
\affiliation{Maryland Quantum Materials Center, Physics Department,\\
University of Maryland, College Park, MD 20742-4111  USA}
\affiliation{Canadian Institute for Advanced Research, Toronto, Ontario M5G 1Z8, Canada}
\author{Nicholas P. Butch}
\affiliation{Maryland Quantum Materials Center, Physics Department,\\
University of Maryland, College Park, MD 20742-4111  USA}
\affiliation{NIST Center for Neutron Research, National Institute of Standards and Technology,\\
Gaithersburg, Maryland 20899, USA}
\author{Steven M. Anlage}
\affiliation{Maryland Quantum Materials Center, Physics Department,\\
University of Maryland, College Park, MD 20742-4111  USA}
\date{February, 2025}

\begin{abstract}
We explore the pairing mechanism and possible topological properties of the unconventional superconductor \ute through their influence on both the inductive and resistive channels using resonant cavity perturbation techniques. We measure a composite surface impedance for a variety of \ute single crystals at microwave frequencies, where the inductive and resistive channels are comparable and can be treated on equal footing. By studying the composite electrodynamics for a large number of resonant modes, we are able to provide a comprehensive picture of the directional dependence of the excitations out of the ground state. We find similar power-law temperature dependencies at low-temperature for both the penetration depth and surface resistance, which in aggregate across modes, are quadratic and isotropic. On the other hand, we find a distinct and systematic anisotropy of the residual loss. The power-law results are inconsistent with a single pair of point nodes on the Fermi surface under weak scattering. We discuss the possibility of topological superconductivity as an interpretation of these isotropic power-laws in conjunction with the anisotropic residual loss.

\end{abstract}

\maketitle

\section{Introduction}
The underlying properties of a superconductor, such as the pairing mechanism, Fermi surface, and topological properties, can imprint themselves on the low-temperature behavior of observable quantities. The electrodynamics of such low-energy excitations out of the superconducting ground state are commonly studied through the inductive channel, such as with the magnetic penetration depth. The resistive superconducting channel, though, is far less commonly studied, especially since it is not accessible to dc or low-frequency measurements. Residual loss, however, is an important feature of exotic superconductors such as \ute[,] and the resistive channel should also be sensitive to the low-energy excitations, though it is more susceptible to other influences. 

The heavy Fermion superconductor \ute has been a hot topic for several years now as it demonstrates a confluence of exotic properties, including evidence of spin-triplet pairing, nodes in the superconducting gap, and topological surface states. During this time, the material has undergone multiple iterations of growth methods, namely chemical vapor transport (CVT) and molten salt flux (MSF). The CVT-grown samples for which superconductivity was first reported in this material had $T_c\sim 1.6$ K \cite{Ran19,Ran21_CVT}. Since then, the transition temperature has been increased to $T_c \lesssim 2$ K in these CVT-grown samples \cite{Rosa22}. Subsequently, very high-quality samples (RRR $\sim100$) with even higher transition temperatures of $T_c\approx 2.1$ K have been produced using a MSF growth method \cite{Sakai22,Aoki24}. The $T_c$ apparently saturates for RRR $\gtrsim 50$, though, and samples with different RRR values can have the same $T_c$ \cite{Rosa22,Sakai22,Aoki24}. There is considerable variation in RRR between samples across and within generations \cite{Svanidze25}.

The upper critical field in \ute has been observed to exceed both the Pauli and orbital limits, suggesting triplet superconductivity \cite{Ran19}. Additionally, the expected temperature dependence of the NMR Knight shift for singlet superconductors is not observed \cite{Ran19,Nakamine19}. In the first generation of CVT samples, a double transition in the specific heat as well as the onset of a non-zero polar Kerr effect at $T_c$ were observed, suggesting time-reversal invariance (TRS) breaking and chiral superconductivity \cite{Hayes21}. Despite this, no evidence of TRS breaking has been seen in the subsequent generations \cite{Ajeesh23,Azari23,Gu25}.

Since the first generation of CVT samples, evidence for nodal superconductivity has been seen in this material using electrodynamic \cite{Metz19,Seokjin21,Ishihara23} and thermodynamic \cite{Ran19,Hayes21,Hayes25} measurements. Experiments to determine the specific nodal configuration have continued for the proceeding generations of samples. Based on the symmetry of the crystal structure, there are certain allowed configurations of the superconducting gap structure, called irreducible representations; however, non-symmetry-imposed nodes are still possible. For the $D_{2h}$ (orthorhombic) point group crystal symmetry of \ute with triplet superconductivity, the possibilities are a full gap ($A_u$) or point nodes along the $a$-, $b$-, $c$-directions ($B_{3u}$, $B_{2u}$, and $B_{1u}$, respectively). For a singlet superconducting state, there could be line nodes instead of point nodes \cite{Blount85,Metz19}. 

There have been many conflicting reports in the literature for the superconducting order parameter with each of the listed irreducible representations being claimed by various authors \cite{Theuss24}. For example, thermal conductivity measurements have been performed by multiple researchers with varying results \cite{Hayes25,Suetsugu24,Hosoi26}. These experimentally deduced single-component order parameter results range from fully-gapped $A_u$ \cite{Suetsugu24,Hosoi26} to to point nodal including $B_{1u}$ \cite{Li25}, $B_{2u}$ \cite{Hayes25,Hayes26}, as well as $B_{3u}$ from STM \cite{WangS25}.

These proposed order parameters have been single-component. For a chiral system, it is possible to have a multi-component order parameter which is a complex combination of two irreducible representations; however, the literature does not support chiral superconductivity in \ute[.] Nonetheless, a multi-component combination of $B_{3u}$ and $A_u$ has been concluded from magnetic screening measurements, albeit with caveats \cite{Ishihara23,Iguchi23,Kamat26}.

While there is no general agreement on the superconducting order parameter, thermodynamic measurements \cite{Hayes25,Hayes26} tend to indicate nodes along the $b$-axis while surface-sensitive measurements \cite{Gu25,WangS25} tend to indicate nodes along the $a$-axis. The possible existence of topological \cite{Roising24,Chris25} or other surface states \cite{Aishwarya23,Gu23} could contribute to the differences between results obtained from surface and bulk measurements of \ute[.] There are some measurements that conclude nodes along the $c$-axis; however, due to the quasi-2D nature of the Fermi surface \cite{Eaton24,Broyles23}, this can be difficult to experimentally realize. 

In this paper, we perform surface impedance and magnetic penetration depth analyses for two MSF-grown \ute single crystals. We perform three separate measurements of one of these samples in orthogonal orientations to thoroughly probe the possible configurations of screening current flow directions, increasing the number of modes studied from 31 in Ref.~\cite{Arthur25} (which only consisted of CVT-grown samples) to 130 modes across all samples, both in this work and Ref.~\cite{Arthur25}. This collection of modes densely samples most of the space of induced-current weights, where only a small region was previously sampled. This reveals features which were previously unknown and provides a substantial level of confidence to our results. We study the low-temperature power-law behavior of the penetration depth and surface resistance and find isotropic, $\sim T^2$ behaviors for both, which is inconsistent with simple point nodes under weak scattering. In contrast, we find a systematic anisotropy of the residual loss, which we discuss in the context of possible surface states.

We will first give an overview of the experiment then discuss the results of low-temperature excitations through both the penetration depth and the surface resistance. We will then present the residual loss at zero-temperature and finally discuss the interpretation of our results as a whole.

\section{Results}

We use microwave cavity perturbation techniques \cite{Waldron67,Sridhar88,Rubin88,Bonn91,Anlage92,Hardy93,Mao95,Trunin98,Ormeno02,Seokjin21} 
to determine the surface impedance of \ute single crystals. These experiments have previously been performed on CVT-grown \ute samples \cite{Seokjin21,Arthur25}. Here we measure MSF-grown samples to provide a more complete picture of the electrodynamics of this exotic material. As in Ref.~\cite{Arthur25}, our measured surface impedance and the quantities which we derive from it are \textit{composite} in the sense that they arise from screening currents flowing in multiple crystallographic directions. We are, however, able to use the characteristic anisotropy of the normal state dc resistivity tensor to estimate the fractions of the induced screening currents which flow in each crystallographic direction, which we call the induced-current weights $w_i\ (i=a,b,c)$. This information is also critical to determine the surface impedance from the underlying resonance data as discussed in Ref.~\cite{Arthur25}. It is clear that high-quality dc resistivity tensor data for MSF-grown samples is necessary to characterize the new MSF samples which we measure. We discuss this independently measured dc resistivity data for MSF samples and the fitting we use to determine the surface impedance in Appendix~\ref{sec:rho_data}.

\begin{figure}	
\hspace*{-0.5cm}\includegraphics[scale=0.23,width=0.5\textwidth]{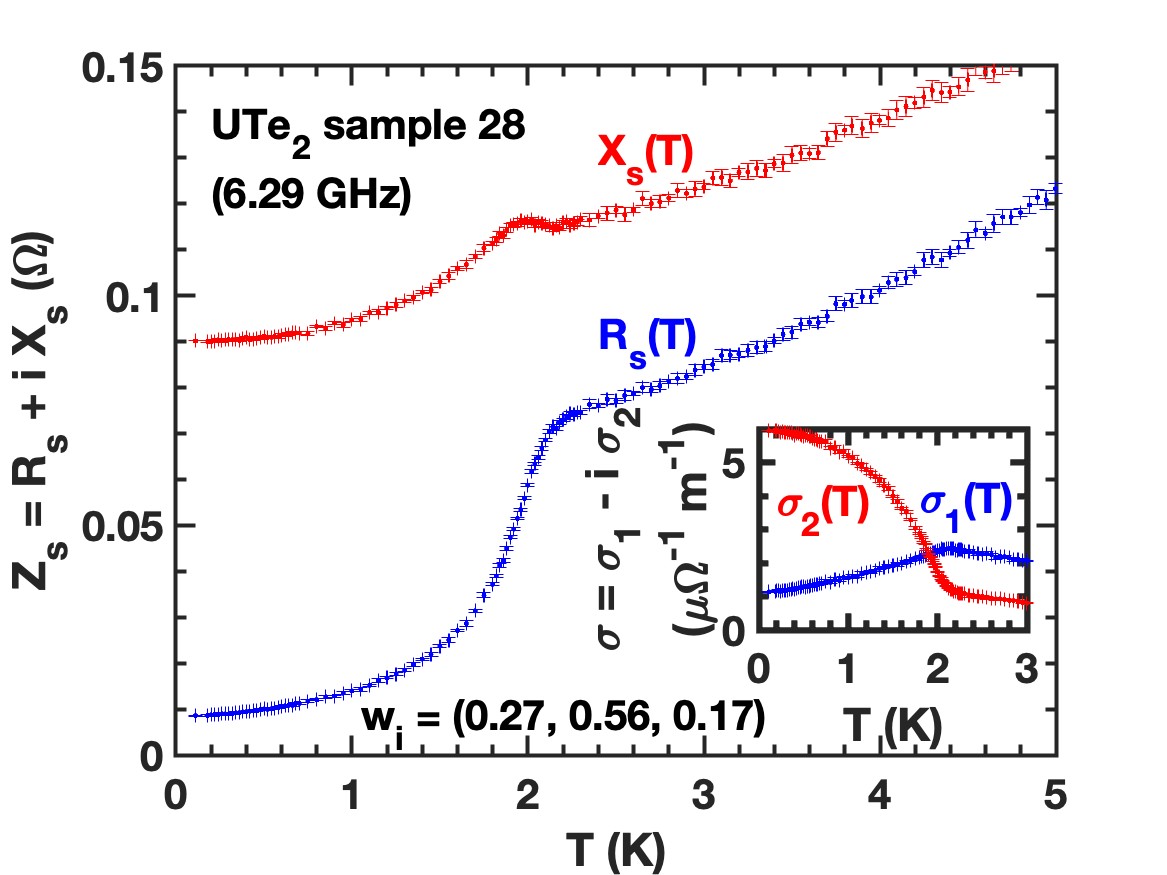}
\caption{Plots vs temperature in the vicinity of the superconducting transition of the surface resistance $R_s(T)$ in blue and surface reactance $X_s(T)$ in red, demonstrating a peak just below $T_c$, for a particular mode with a MSF-grown \ute sample. The complex conductivity is similarly plotted in the inset with real and imaginary parts shown in blue and red, respectively.}
\label{fig:Zs_vs_T}
\end{figure}	

An example of the surface impedance $Z_s(T) = R_s(T) + i X_s(T)$ for a particular mode of one the MSF samples we study is shown in Fig.~\ref{fig:Zs_vs_T}. As with the CVT samples discussed in Refs.~\cite{Seokjin21,Arthur25}, a reactance peak is observed in $X_s(T)$ just below the critical temperature $T_c$. As shown in Fig.~\ref{fig:Zs_vs_T}, while the surface resistance $R_s(T)$ rapidly drops as the temperature crosses below $T_c$, the surface reactance $X_s(T)$ increases first into this peak before eventually decreasing. A reactance peak is broadly observed to some degree in all of our data for both CVT and MSF samples. This effect is indicative of the competition between the normal and superfluid screening responses (which act in quadrature) just below $T_c$, where the superfluid is initially less effective at screening than the normal fluid. These peaks are a sign of long quasiparticle mean free paths \cite{Hein01,Ormeno02,Ormeno06,Baker09,Baker10,Ghigo18}. 
Note that this differs from the coherence peak which occurs just below $T_c$ in the real part of the complex conductivity $\sigma_1(T)$ for conventional superconductors. The complex conductivity $\sigma(T) = \sigma_1(T) - i \sigma_2(T)$, which is determined from $Z_s(T)$ using the local limit relation $Z_s = \sqrt{\frac{i\mu_0\omega}{\sigma}}$, is shown in the inset of Fig.~\ref{fig:Zs_vs_T}. Below $T_c$, the real part, which is a measure of the loss, shows an immediate monotonic decrease while the imaginary part, which represents the superfluid response in the superconducting state, increases monotonically. There are some `exceptional' modes discussed in Appendix~\ref{sec:sigma} for which $\sigma_1(T)$ monotonically increases below $T_c$ or demonstrates a peak; however, these are also not `coherence peaks'.

To investigate excitations out of the ground state and possible nodal structures of \ute[,] we carefully study the temperature dependence of the magnetic penetration depth, which we determine from the complex conductivity using $\lambda(T)=1/\sqrt{\mu_0\omega\sigma_2(T)}$. For low temperatures ($T<T_c/3$) we fit our penetration depth data to a power-law form given by,
\begin{equation}
    \frac{\lambda(T)-\lambda_0}{\lambda_0} = \eta \left(\frac{T}{T_c}\right)^{\alpha},
    \label{eq:lambda_pow_law}
\end{equation}
which is predicted for nodal superconductors \cite{Gross86,Klemm88}. From these fits, we extract the zero-temperature penetration depth $\lambda_0$, power-law exponent $\alpha$, and power-law coefficient $\eta$ as parameters for each measured mode in each measurement of a sample. The parameters $\alpha$ and $\eta$ have theoretical predictions for various nodal configurations on the Fermi surface \cite{Gross86,Klemm88}, to which we will compare our results below.

Since our measurement is sensitive to the microwave superconducting loss, we also carefully study the temperature dependence of the surface resistance. We observe a power-law temperature dependence at low-temperature,
\begin{equation}
   \frac{R_s(T)-R_0}{R_0} = \eta_R \left(\frac{T}{T_c}\right)^{\alpha_R},
   \label{eq:Rs_pow_law}
\end{equation}
which we write in an analogous form to Eq.~\ref{eq:lambda_pow_law}. We determine the zero-temperature surface resistance $R_0$, power-law exponent $\alpha_R$, and power-law coefficient $\eta_R$ as fitting parameters for each measured mode in each measurement of a sample. Unlike the power-law temperature dependences of $\lambda$, those of $R_s$ do not have such well established interpretations; however, they very likely arise from the same underlying excitations out of the ground state, and $R_0$ is an important measure of the residual loss at zero-temperature. Our measurements provide an important insight into the superconducting loss as a tool to study the excitations out of the ground state.

In Ref.~\cite{Arthur25}, the penetration depth and surface resistance were found to have comparable values for $\alpha$ and $\alpha_R$. Across modes, these demonstrated isotropic values near or less than $2$. The residual loss, on the other hand, demonstrated a distinct anisotropy with larger loss for a certain region of the induced-current weight parameter space, which corresponded to the `exceptional' modes demonstrating a monotonic increase in $\sigma_1(T)$ with decreasing temperature. Much of the induced-current weight parameter space was un-probed, though, and these results were only for CVT-grown samples. To expand upon these results in this paper, we measure two MSF-grown \ute samples. One of these samples (23 B) was needle-shaped along the $a$-direction, which is how crystals typically form for this growth method, while the other sample (28) has comparable dimensions along each of the crystallographic axes, similarly to samples of the CVT growth methods. Additionally, sample 28 has facets along the crystallographic axes, allowing it to be oriented in our measurements. We measure sample 28 three separate times under orthogonal orientations to probe as many distributions of the induced screening currents as possible. See Appendix~\ref{sec:materials} for information about the growth of the samples discussed here.

\subsection{Low-temperature excitations}

\begin{figure}[!ht]	
\begin{subfigure}[b]{0.4\textwidth}
\caption{}
\hspace*{-2cm}
\includegraphics[scale=0.23,width=1.0\textwidth]{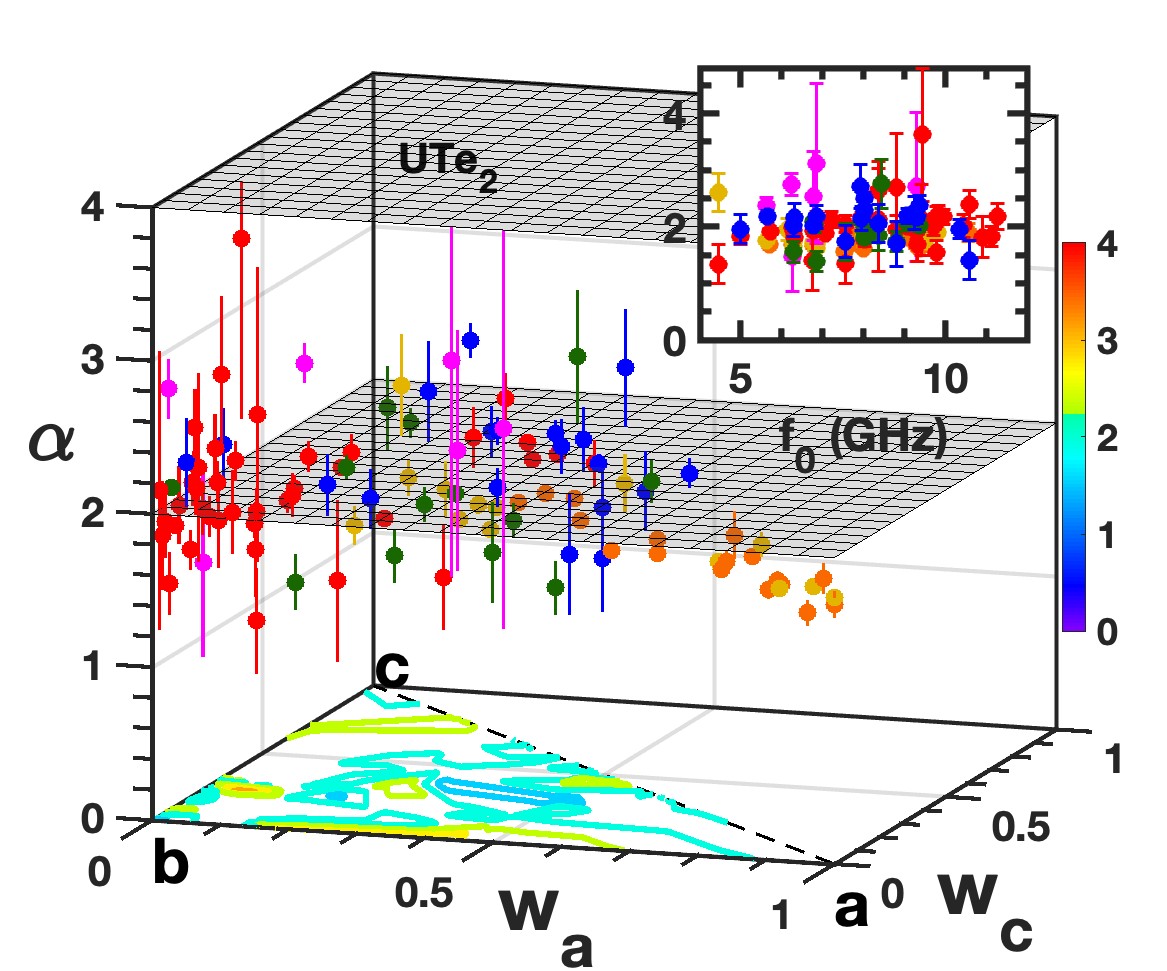}
\label{fig:alpha_vs_w}
\end{subfigure}
\begin{subfigure}[b]{0.4\textwidth}
\caption{}
\hspace*{-2.3cm}
\includegraphics[scale=0.23,width=1.0\textwidth]{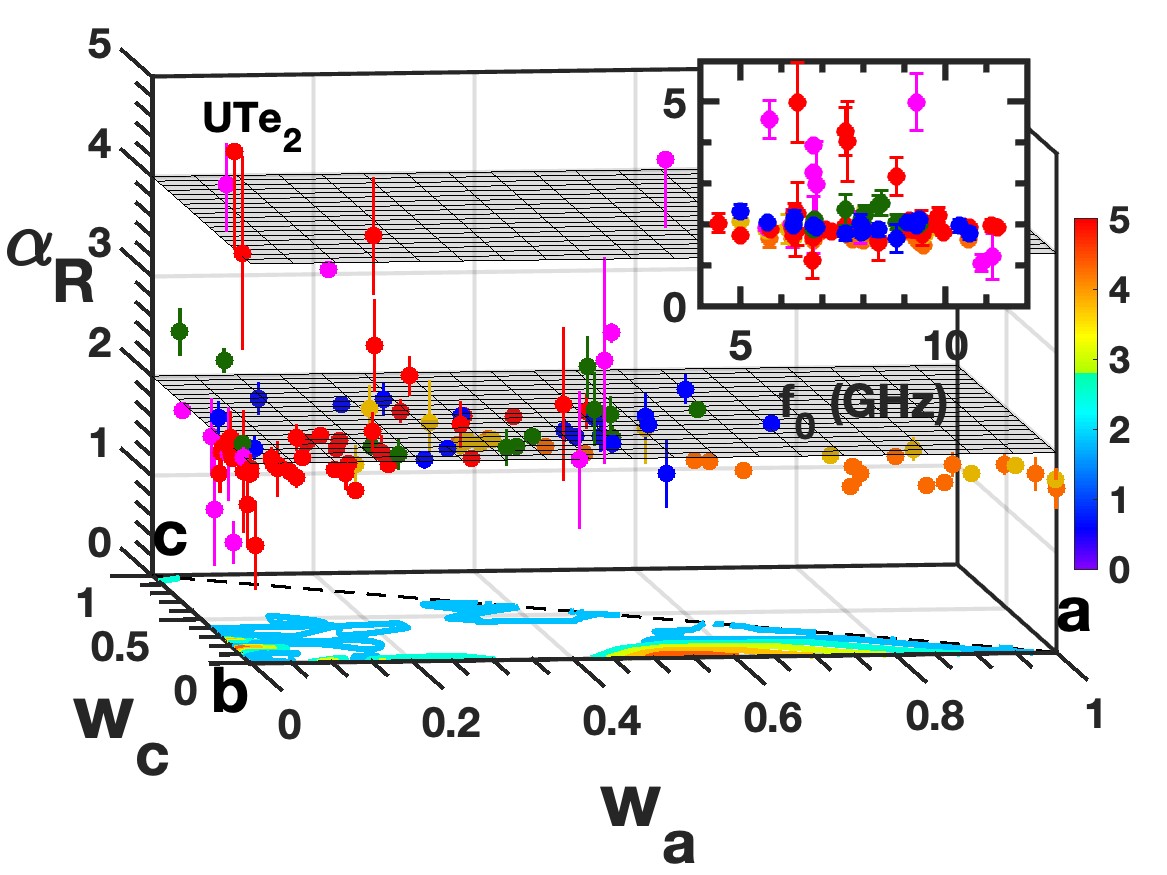}
\label{fig:alpha_R_vs_w}
\end{subfigure}
\caption{Plots of (a) penetration depth power-law exponent $\alpha$ and (b)  surface resistance power-law exponent $\alpha_R$ vs current-flow contribution of each crystallographic axis (and vs mode frequency in the insets). The simple $p$-wave predictions for the \underline{penetration depth} of $\alpha=2$ with currents parallel to the point nodes and $\alpha=4$ for currents perpendicular to the point nodes are indicated as horizontal planes. 
Data for two CVT-grown \ute samples B39 and B40 is shown in orange and yellow, respectively. For the MSF-grown \ute samples, data for sample 23B is shown in magenta while data for sample 28 is shown in red, green, and blue for the cases of the $a$-, $b$-, and $c$-axes aligned with axis of the `hot finger', respectively. Contours of the data are also included in the $w_a$-$w_c$ plane with colors (indicated in the colorbars) corresponding to the value of the quantity being plotted.}
\label{fig:w_dep}
\end{figure}

We will now discuss the results of the power-law exponent $\alpha$ from fittings of the composite penetration depth data $\lambda(T)$ to the power-law form given by Eq.~\ref{eq:lambda_pow_law} for multiple \ute samples. We will additionally discuss the results of the power-law exponent $\alpha_R$ from fittings of the composite surface resistance data $R_s(T)$ to the power-law form given by Eq.~\ref{eq:Rs_pow_law}. 
As with the CVT samples discussed in Ref.~\cite{Arthur25}, we find no systematic dependence of $\alpha$ or $\alpha_R$ on the crystallographic contributions as shown in Fig.~\ref{fig:w_dep}. The error bars shown in Fig.~\ref{fig:w_dep} represent the uncertainty in the determination of the fit parameters corresponding to a $3\%$ degradation of the fit quality. 
Contours of the $\alpha$ and $\alpha_R$ data are shown in the $w_a$-$w_c$ plane with colors (indicated in the colorbars) corresponding to the value of the quantity being plotted. The contours help to visualize location in the induced current weighting space of values of $\alpha$ and $\alpha_R$ which are much different from the rest. The large values of $\alpha$ and $\alpha_R$ for a few modes can be seen as small contour loops with warm color tones. Unlike the residual loss, which will be discussed below, these modes do not appear to form a clear systematic dependence on the induced-current weights, rather they are isolated without consistently similar values in neighboring modes in terms of $w_i$. For most regions of the induced current weighting space, though, the contours of $\alpha$ and $\alpha_R$ have intermediate values, indicating that the data fluctuates around these intermediate values without a particular systematic dependence on the induced current weights.
The $\alpha=2$ and $\alpha=4$ predictions for currents parallel and perpendicular, respectively, to point nodes \cite{Klemm88} are shown as horizontal planes in Fig.~\ref{fig:alpha_vs_w}. For reference, Fig.~\ref{fig:alpha_R_vs_w} also includes these predictions for the \underline{penetration depth}. There is no reason \textit{a priori}, though, to expect that the power-law exponent for the surface resistance should correspond to these values; however, we include these values as references since we find empirically that $\alpha$ and $\alpha_R$ have such similar values. 

\begin{table*}
    \begin{center}
    \caption{A comparison of several composite and axis-resolved electrodynamic properties for various \ute samples.}
    \hspace*{-1cm}\begin{tabular}{ | c | c | c | c | c | c | c | c | }
     Batch/ citation & Growth & $T_c$ & $\alpha$ & $\eta$ & $\lambda_0$  & $\alpha_R$ & $R_0/\omega^2$ \\ 
     & method & (K) &  &  & $(\mu m)$ &  & $(\Omega\ ps^2)$ \\  
     \hline
     B39 & CVT & $1.96$ & $1.70 \pm 0.06$ & $0.52 \pm 0.17$ & $0.64 - 2.01$ & $1.73 \pm 0.08$ & $0.02 - 5.41$ \\ 
     B40 & CVT & $1.96$ & $1.72 \pm 0.06$ & $0.56 \pm 0.11$ & $1.00 - 2.42$ & $1.79 \pm 0.06$ & $0.05 - 22.39$ \\
     23B & MSF & $2.25$ & $2.34 \pm 0.49$ & $2.21 \pm 3.68$ & $0.62 - 2.00$ & $3.39 \pm 0.84$ & $0.00 - 6.59$ \\
     28 a & MSF & $2.13$ & $1.36 \pm 0.47$ & $0.47 \pm 0.12$ & $0.80 - 3.81$ & $1.88 \pm 0.11$ & $0.00 - 26.34$ \\
     28 b & MSF & $2.13$ & $2.00 \pm 0.17$ & $0.45 \pm 0.14$ & $1.03 - 2.81$ & $2.05 \pm 0.13$ & $0.15 - 9.21$ \\
     28 c & MSF & $2.13$ & $2.16 \pm 0.19$ & $0.48 \pm 0.21$ & $1.40 - 3.91$ & $1.98 \pm 0.12$ & $2.85 - 18.48$ \\
     28 (aggregate) & MSF & $2.13$ & $1.44 \pm 0.50$ & $0.47 \pm 0.12$ & $0.80 - 3.91$ & $1.90 \pm 0.12$ & $0.00 - 26.34$ \\
     \hline
     S. Bae (UMD) \cite{Seokjin21} & CVT & $1.57$ & $2.11$ &  & $0.791$ &  & $3.0$ \\ 
     \hline
     K. Ishihara (Tokyo) \cite{Ishihara23} & CVT & $1.65$ & ($2.01$,$1.90$,$1.60$) &  &  &  & \\ 
      & CVT & $1.75$ & ($2.08$,$2.06$,$1.84$) &  &  &  & \\
      & MSF & $2.1$ & ($2.07$,$2.16$,$2.04$) &  & ($1.42$,$0.710$,$2.75$) &  &
     \label{tab:lambda_and_loss}
    \end{tabular}
    \end{center}

\end{table*}

Since the composite values of $\alpha$ and $\alpha_R$ (as well as $\eta$) show no systematic dependence on estimates of the axis weights, we present average values of these quantities for each sample as well as the standard deviation among the modes as a measure of the uncertainty of these averages in Table \ref{tab:lambda_and_loss}. Considering these uncertainties, $\alpha$ and $\alpha_R$ are practically equivalent; although, for sample 23B, the power-law exponent averages are considerably different, but they do have very large uncertainties. For other quantities in this table which show a systematic dependence on the crystallographic responses, we present the range of values observed for all modes measured. We compare our results with those of other researchers in Table~\ref{tab:lambda_and_loss}. The results from Ref.~\cite{Seokjin21} in Table~\ref{tab:lambda_and_loss} are from composite $\lambda(T)$ data for a single resonant mode ($f_0=11.26$ GHz) of the same cavity we utilize. While the results of Ref.~\cite{Ishihara23} presented in Table~\ref{tab:lambda_and_loss} are axis-resolved, these dc tunnel-diode oscillator screening measurements are not sensitive to loss, such as the surface resistance or $\sigma_1$.

Measures of the microwave superconducting loss, such as the surface resistance $R_s$ and real part of the complex conductivity $\sigma_1$, are more difficult to interpret than inductive quantities, such as the penetration depth. 
Though there is a predicted intrinsic value for the residual superconducting loss in the case of line-nodes \cite{Hirsch93}, such a universal value cannot be established for a general nodal structure. Additionally, Ref.~\cite{Hirsch93} shows that in addition to the number of quasiparticles excited, the low-temperature behavior of the loss depends on the quasiparticle scattering rate, the scattering phase shift, and the magnetic penetration depth. Despite this, our observation of simple power-law behavior of $R_s(T)$ at low temperatures motivates its efficacy as a tool to study excitations out of the ground state. Furthermore, the similarity between the penetration depth and surface resistance power-laws suggests that these are probing a common phenomenon.

\subsection{Zero-temperature loss}

\begin{figure}[!ht]	
\hspace*{-0.5cm}\includegraphics[scale=0.23,width=0.6\textwidth]{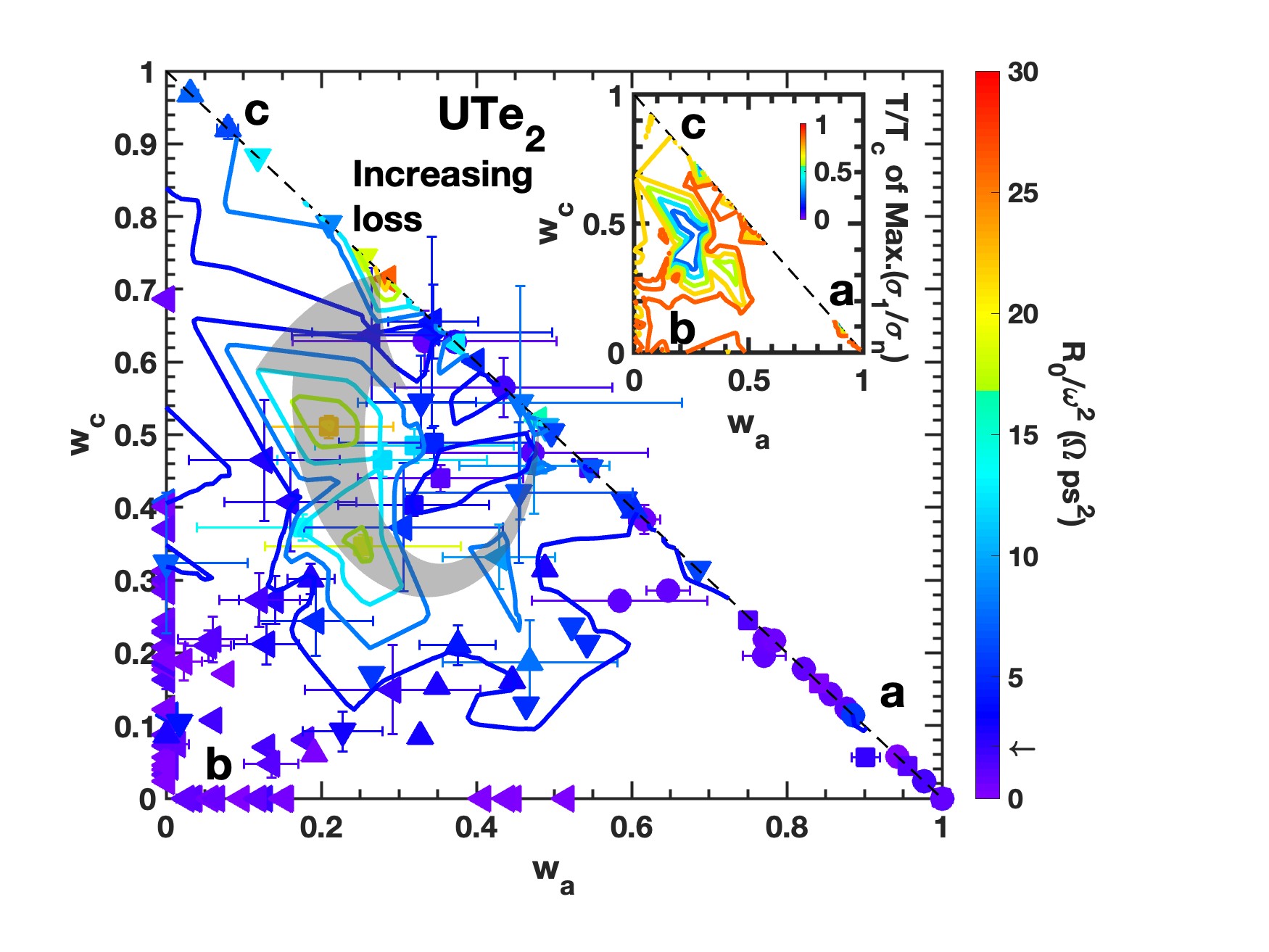}
\caption{Plot of nominally frequency independent zero-temperature loss $R_0/\omega^2$ vs current-flow contribution of each crystallographic axis. 
Data for two CVT-grown \ute samples B39 and B40 is shown as circles and squares, respectively. For the MSF-grown \ute samples, data for sample 23B is shown as left triangles while data for sample 28 is shown as left, up, and down for the cases of the $a$-, $b$-, and $c$-axes aligned with axis of the `hot finger', respectively. The colors indicate the value of $R_0/\omega^2$ in units of $\Omega$ ps$^2$. The large, gray arrow indicates the direction of increasing loss along the arc of large residual loss. Contours of the $R_0/\omega^2$ data are also included with the same color scale (indicated in the main colorbar) corresponding to the value of $R_0/\omega^2$. The estimate of the universal $d$-wave loss is also marked with an arrow in this colorbar. 
The inset shows the contours of the fraction of $T_c$ at which $\sigma_1(T)/\sigma_n$ has its maximum below $T_c$ indicating the exceptional modes (which are blue according to the colorbar in this inset).}
\label{fig:Rs_0_vs_w}
\end{figure}	

In a BCS superconductor, the finite temperature loss at frequencies below the gap is expected to scale quadratically with frequency, $R_s(T)\sim \omega^2$, in the absence of a residual loss mechanism. Additionally, a roughly quadratic frequency dependence has been observed in the zero-temperature surface resistance in CeCoIn$_5$ \cite{Ormeno02}. For line-nodal superconductors, the universal residual surface resistance also scales quadratically with frequency, namely  $R_0^{line-nodal} = \mu_0\omega\lambda_L \frac{\hbar\omega}{2\Delta_{\text{max.}}}$ \cite{Hirsch93}. To provide a benchmark with which to compare our residual loss values, we estimate $R_0^{line-nodal}\approx 8.4$ m$\Omega$ at $10$ GHz (corresponding to $R_0^{line-nodal}/\omega^2 = 2.1\ \Omega\  ps^2$) for \ute[,] assuming $\Delta_{\text{max.}}=0.29$ meV and $\lambda_L = 1.5$ $\mu$m. Aside from the residual loss expected from nodal quasiparticles, Andreev (or Majorana, etc.) bound states can also contribute to loss in nodal superconductors. In chiral $p$-wave superconductors, bound states are also predicted to contribute to the residual loss \cite{Mats99,Bakur18}.



We use $R_0/\omega^2$ as a nominally frequency independent measure of the residual loss, where $R_0$ is the zero-temperature surface resistance determined from fits to Eq.~\ref{eq:Rs_pow_law}. This quantity relies almost directly on raw data as opposed to other frequency independent measures of the loss, such as the real part of the complex conductivity, which involve more assumptions and processing. We do not observe a clear, systematic frequency dependence of $R_0/\omega^2$ for our data. We plot $R_0/\omega^2$ vs the current-induced weights in a two-dimensional, top-down view in Fig.~\ref{fig:Rs_0_vs_w} for which the value of $R_0/\omega^2$ is indicated by the color of each data point. The error bars shown in Fig.~\ref{fig:Rs_0_vs_w} represent the uncertainties in the determination of $w_i$ as fit parameters in Eq.~\ref{eq:geo_mean_comp} corresponding to a $3\%$ degradation of the fit quality. To further accentuate regions of large residual loss in the parameter space of induced-current weights, we also include contours of the $R_0/\omega^2$ data. The estimate of the line-nodal residual loss is indicated with an arrow on the $R_0/\omega^2$ colorbar. A systematic dependence of the residual loss on the induced-current weights can be seen in Fig.~\ref{fig:Rs_0_vs_w} which peaks for an arc of current flow directions indicated by the large, gray arrow with endpoints having no current in the $b$-direction. The $a$-/ $c$-axis current contributions at these endpoints are about $30/70\%$ and $50/50\%$. We observe the largest frequency-independent loss at the former endpoint. The (relatively) large values of the frequency-independent loss on this arc tend to increase along the arc from the latter toward the former endpoints, as indicated by the widening gray arrow in Fig.~\ref{fig:Rs_0_vs_w}.
The zenith of this arc has a contribution of $b$-direction of about $40\%$. The ranges of values for $R_0/\omega^2$ across various samples are also included in the summary Table~\ref{tab:lambda_and_loss}.

The contours of the fraction of $T_c$ at which $\sigma_1(T)/\sigma_n$ has its maximum below $T_c$ are shown in the inset of Fig.~\ref{fig:Rs_0_vs_w} for comparison with the region of large residual loss in the parameter space of induced-current weights. This region of large residual loss does not appear to be correlated with extreme values of the other power-law parameters of $\lambda(T)$ and $R_s(T)$, such as $\alpha$, which occur for a few modes. Furthermore, the modes nearby these few modes in the induced-current weights do not show increased values of these other power-law quantities as they do for $R_0/\omega^2$. For both $R_0/\omega^2$ and $T/T_c$ of maximum $\sigma_1/\sigma_n$ below $T_c$, the modes demonstrating large loss or low-temperature enhancement of $\sigma_1$ are contained in, what is apparently, the same distinct region of the induced current weight phase space, for which nearby modes exhibit similar behavior.

It is unlikely that two-level system (TLS) loss plays a role in the residual loss described here. See Appendix~\ref{sec:TLS} for further discussion of the possibility of TLS loss in this system.

\section{Discussion}
We have thoroughly probed the space of possible current flow directions through 130 resonant modes across the \ute samples we have studied. This fills out the $w_i$ parameter space as can be seen in the top-down view of this space in Fig.~\ref{fig:Rs_0_vs_w}. Despite this, there is no systematic variation with respect to the induced-current weights of the penetration depth or surface resistance power-law exponents shown in Fig.~\ref{fig:w_dep} to indicate the direction of a nodal symmetry under weak scattering. In the simple point nodal ($p$-wave) case with weak scattering, we would expect $\alpha=2$ for currents along the direction of the nodes while $\alpha=4$ would be predicted for currents in the plane perpendicular to the nodal symmetry axis \cite{Klemm88}. We would expect this to manifest as a distinct anisotropy between a few modes consistent with the parallel-to-nodes results and many more modes consistent with the perpendicular-to-nodes results since the space of current flow directions perpendicular to the nodal symmetry axis is higher dimensional than that along it. Similarly, for a line node under weak scattering, a characteristic anisotropy would be expected, but with $\alpha=3$ along the symmetry axis and $\alpha=1$ in the perpendicular direction (currents through the nodes) \cite{Klemm88}. In this case, the higher power-law exponent would only correspond to a few modes. Our data, apparently, demonstrates excitations out of the ground state on the order of currents near points nodes for all current directions, without any apparent anti-nodes. This is inconsistent with the simple $p$- and $d$-wave cases just discussed. The case of line nodes is, additionally, particularly unlikely since the residual loss values we observe are less than our estimate of the universal $d$-wave prediction for many modes. The case of strong (unitary) scattering, where the excitations at the nodes become averaged across the Fermi surface, could produce this behavior. For both the $p$- and $d$-wave cases under unitary scattering, $\alpha\approx 2$ is predicted \cite{Klemm88}. However, the high RRR values of the crystals and consistent appearance of the $X_s(T)$ peaks below $T_c$ argue against this interpretation. It has been proposed that multiple pairs of point nodes slightly off of one or more of the high symmetry axes could produce such an isotropic $\alpha\approx 2$ result \cite{Ishihara23,Chris25}. Similar, nearly isotropic power-laws with $\alpha = 1.9-2$ in PrOs$_4$Sb$_{12}$ were attributed to multi-domain orientations of a point nodal order parameter \cite{Chia03}.

A possible mechanism which is qualitatively consistent with our isotropic power-law exponent results shown in Fig.~\ref{fig:w_dep} as well as the anisotropic residual loss results shown in Fig.~\ref{fig:Rs_0_vs_w} (indicated by the large, gray arrow) is the presence of topological Weyl nodes. Such a scenario would require broken time reversal symmetry, though, which is not consistently supported by the literature. There are multiple conjectures, however, which seek to devise a means by which such a state could still be possible, such as a combined Kondo/spin-liquid model \cite{Cole24} or a mixed order parameter arising from nearly degenerate irreducible representations \cite{Choi24}. 

For a topological Weyl superconductor, it is predicted that the penetration depth power-law exponent would have an isotropic value of $\alpha=2$ \cite{Foster21}. Qualitatively, this matches our results for $\alpha$ shown in Fig.~\ref{fig:alpha_vs_w} as well as those reported in Refs.~\cite{Seokjin21,Ishihara23}. The observed values of $\alpha$ are close to this prediction in aggregate, though there are several observed values which depart from this prediction (though not systematically). Most notably, the isotropy of $\alpha$ is a striking feature which is difficult to explain simply with nodes. This topological model is compelling for its simplicity, though multiple nodes may still be present in the topological case.
Refs.~\cite{Roising24,Chris25} discuss theoretical predictions and consequences of Majorana-Dirac surface states in \ute[.] Christiansen \etal conclude that for $B_{2u}$ and $B_{3u}$ order parameters, several non-symmetry-imposed nodes would be necessary in addition to the symmetry-imposed nodes, which is conceivable considering the isotropic $\sim T^2$ power-laws that have been observed for the penetration depth.

The second feature of a topological Weyl superconductor is that the loss is predicted to have a characteristic anisotropy, corresponding to larger loss for electric fields polarized in the direction of the Weyl nodes. This loss arises from the absorption of a photon from the gapless 2D Majorana fermion surface fluid to the bulk. This loss mechanism is predicted to contribute to the zero-temperature residual loss and to be robust to disorder \cite{Foster21}. 
The arc we have observed in the space of induced-current weights (illustrated by the large, gray arrow in Fig.~\ref{fig:Rs_0_vs_w}) could possibly indicate the presence of Weyl nodes along it if such a mechanism were active in the material. 
Such an effect is not necessarily unique to this theory; the concept of a surface to bulk photon absorption as a loss mechanism could arise from other topological scenarios, potentially one which does not require broken time reversal symmetry. Ref.~\cite{Gu25} find a zero-energy Andreev conductance peak on the $(0\bar{1}1)$ surface, which they conclude is consistent with a TRS preserving topological state.


\section{Conclusions}
In conclusion, we have analyzed the electrodynamic properties of both CVT- and MSF-grown \ute samples for $130$ microwave-frequency modes at low temperatures using the resonant cavity perturbation measurements described in Ref.~\cite{Arthur25}. By studying so many modes with diverse structures of surface currents across multiple samples, we thoroughly probe the possible anisotropy of the properties of \ute[.] Unexpectedly, we find similar power-law temperature dependencies at low-temperature for both the penetration depth and surface resistance, which in aggregate across modes, are quadratic. We do not observe any particular dependence of these power-law exponents on our estimates of the directions of screening current flows. We do, however, find a distinct anisotropy of the residual loss, which forms an arc in the space of induced current directions. Our power-law results are inconsistent with a single pair of point nodes under weak scattering. These isotropic power-laws in conjunction with the anisotropic residual loss are more consistent with topological Weyl superconductivity, though this would require TRS breaking. Alternatively, it is conceivable that a similar loss mechanism involving a surface to bulk photon absorption could occur for some other topological or surface state.

\section{Acknowledgments}
This work was supported by the National Science Foundation grant no. DMR2004386 and ARO/FSDL grant no. W911NF-24-1-0153 (microwave experiments), the U.S. Department of Energy award no. DE-SC-0019154 (sample characterization), the Gordon and Betty Moore Foundation’s EPiQS Initiative through grant no. GBMF9071 (materials synthesis), the National Institute of Standards and Technology, and the Maryland Quantum Materials Center. GSS was supported by NSF DMR2514293. NPB was supported by NIST. SRS acknowledges support from the National Institute of Standards and Technology Cooperative Agreement 70NANB17H301.

\section{Data availability}
The data that supports the findings of this article is openly available. 

\section{Competing interests}
The authors declare no competing interests.

\section{Author contribution}
AC-J performed the microwave cavity perturbation experiments and the corresponding analyses discussed in this paper under the supervision of SMA. AC-J wrote this paper including all text, tables, and figures under the supervision of SMA, JP, and NPB. SRS grew the CVT \ute samples. AC-J prepared these samples for the microwave cavity perturbation measurements. GSS grew the MSF \ute samples and prepared them for the microwave cavity perturbation measurements under the supervision of NPB. JZD and GSS prepared MSF-grown \ute samples for axis-resolved, bar geometry electrical transport measurements, and JZD performed these measurements under the supervision of JP.

\appendix

\section{Methods}
\label{sec:exp_details}
In this appendix, we will discuss the growth methods, preparation, and configurations of the samples used in these experiments (Appendix~\ref{sec:materials}). We will also discuss the anisotropic dc resistivity tensor data used to perform the normal state fittings to determine the surface impedance (Appendix~\ref{sec:rho_data}).

\subsection{Materials}
\label{sec:materials}
The \ute samples B39 and B40 discussed here and in Ref.~\cite{Arthur25} were grown using the CVT technique \cite{Ran21_CVT,Rosa22}. The \ute samples 23B and 28 discussed here were grown using the MSF technique \cite{Sakai22,Aoki24}.

The CVT-grown samples were prepared with a coarse polishing (to shape the crystals) in air followed by a fine polishing (to remove surface oxides) in a reduced oxygen ($\le 0.01 \%$ O$_2$ by volume) environment with no particular orientation of the final surfaces, as discussed in Ref.~\cite{Arthur25}. These samples were then coated in Apiezon N-grease before being returned to the ambient atmosphere. The final side lengths of these crystals were roughly $0.5-1$ mm. We were able to determine the orientations of the MSF-grown samples by their crystal facets and an understanding of the morphology of this growth method. We used wooden picks to scrap these crystal facets while the samples were submerged in acetone in an attempt to remove the oxide layers. These samples were then dried and encased in N-grease.

The CVT-grown samples were mounted on the tip of the sapphire `hot finger' with no particular orientation of the crystallographic axes, as discussed in Ref.~\cite{Arthur25}. Sample 23B was a needle-shaped sample (as is typical of this growth method) with a length of about $1.3$ mm and a size of about $0.1$ mm in the $bc$-plane cross-section. This sample was mounted on the side of the sapphire `hot finger' rather than on its tip. Sample 28 was a bulk, roughly rectangular shaped sample with well-defined $[1 0 0]$, $[0 1 0]$, and $[0 0 1]$ surfaces rather than a needle. Its side lengths were roughly $1.4$, $0.8$, and $0.5$ mm along the $a$-, $b$-, and $c$-directions, respectively. We performed three separate measurements of this sample on the tip of the `hot finger' with the $[1 0 0]$, $[0 1 0]$, and $[0 0 1]$ surfaces aligned with the axis of the `hot finger' in turn.

\subsection{Impedance fitting and anisotropic resistivity data}
\label{sec:rho_data}
The surface impedance $Z_s = R_s +i X_s$ is given by $R_s = \frac{G}{Q_{\text{sample}}}$ and $X_s = -2G\frac{\Delta f_{0,\text{sample}}}{f_{0,\text{tot.}}(T_0)} + X_0$, where the subscripts `sample' and `tot.' denote the resonant responses of the sample alone and the whole sample and resonator system, respectively \cite{Mao95,Seokjin21}. We determine the surface impedance for each mode by fitting this resonance data to,
\begin{equation}
    \sqrt{R_s(T) X_s(T)} = \sum_{i=a,b,c} w_i \sqrt{\mu_0\omega\rho_i(T)/2},
    \label{eq:geo_mean_comp}
\end{equation}
to determine the geometry factor $G$ and minimum temperature reactance $X_0$ as well as estimate the direction of induced currents described by the induced-current weights $w_i$, where $i=a,b,c$ indicates the crystallographic direction \cite{Arthur25}. This fitting relies on high-quality measurements of the dc resistivity tensor $\rho_i$. For the CVT-grown samples B39 and B40 described in Ref.~\cite{Arthur25}, Montgomery geometry resistivity data from Ref~\cite{Yunsuk22} was used in this fit. We have measured bar geometry resistivity data for MSF \ute samples to describe the MSF-grown samples 23B and 28. The $a$ and $b$ bars were cut from another sample from batch 28 while the $c$ bar was cut from a different MSF batch with a similar RRR. We measured $\rho_a$ and $\rho_b$ for two pairs of samples. Since the bulky MSF samples (such as sample 28) tend to have lower RRR values than those of the needle-shaped ones, we have taken the lower RRR measurements of $\rho_a$ and $\rho_b$ to be most representative of our sample 28 data. The RRR values for the $\rho_a$, $\rho_b$, and $\rho_c$ data which we use are $145$, $53$, and $14$, respectively. 

\begin{figure*}[]  
\hspace*{-0.5cm}\includegraphics[scale=0.23,width=1.0\textwidth]{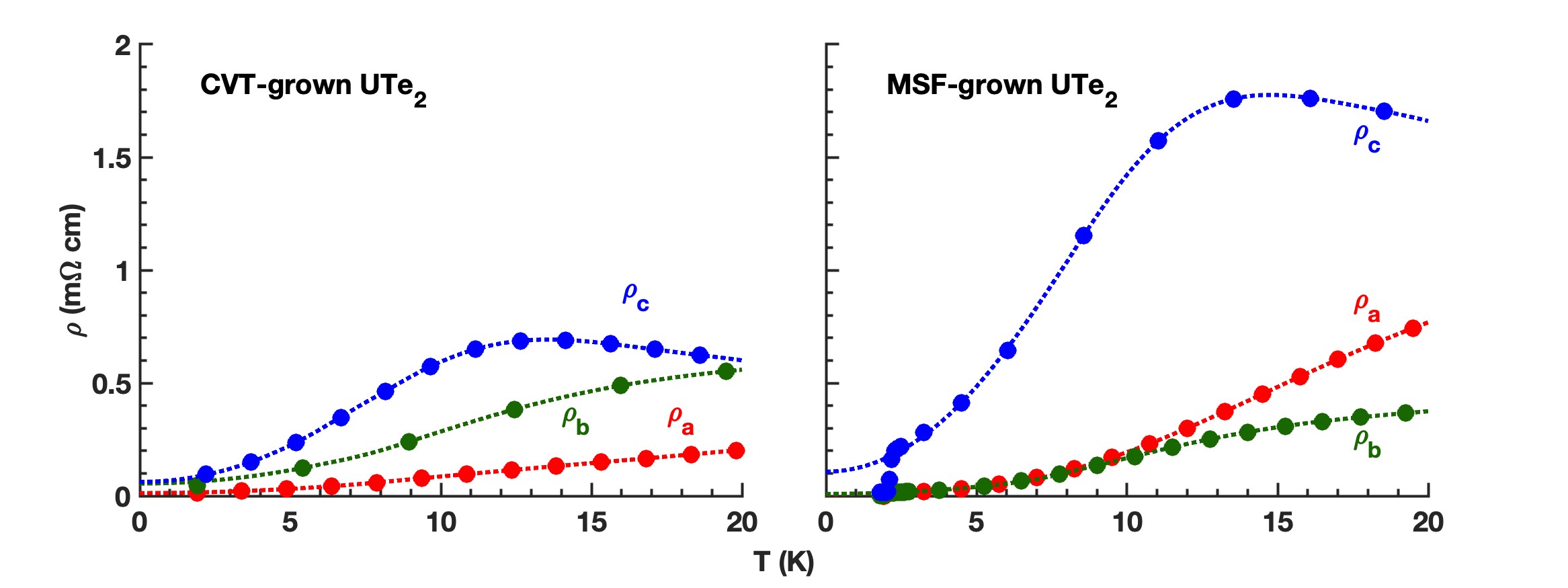}
\caption{Resistivity tensor (decimated) data for two growth methods of \ute samples. The tensor components $\rho_a$, $\rho_b$, and $\rho_c$ are shown in red, green, and blue, respectively. Data for the CVT growth method from Ref.~\cite{Yunsuk22} is shown in the left panel while data for the MSF growth method is shown in the right panel. Our fits (which are quadratic-in-temperature up to $6$ K) to each of these resistivity data sets are shown as dotted lines of the corresponding colors.}
\label{fig:UTe2_rho}
\end{figure*}

At low-temperatures, each component of this MSF resistivity data follows a quadratic power-law $\rho_{i}(T) = A^{(i)} + B^{(i)} T^2$ as was observed for CVT resistivity data \cite{Yunsuk22}. We model each component of our MSF resistivity data with a piecewise polynomial which crosses over between this quadratic power-law below $6$ K and a fifth order polynomial $\rho_{i}(T) = \sum_{n=0}^5 C_n^{(i)}T^n$ from $8-20$ K as described in Ref.~\cite{Arthur25}. A summary of the parameters used in this model is given in Table~\ref{tab:UTe2_MSF_rho_poly}. We compare our MSF resistivity data to the CVT resistivity data from Ref.~\cite{Yunsuk22} in Fig.~\ref{fig:UTe2_rho}. Note that the dc resistivity tensor has been previously measured for MSF-grown \ute microstructures fabricated with a focused ion beam (FIB) method \cite{Zhang25}. Zhang, \etal found that the dc resistivity components for the $a$- and $b$- directions were comparable, which agrees with our resistivity measurements; however, they find a very large anisotropy between the $c$-direction and the $a$- and $b$- directions ($\sim 50$ fold as the temperature approaches zero while we find it to be $\sim 10$ fold). Additionally, the hierarchy of the dc resistivity tensor components for MSF samples in our measurement agrees with that of Ref.~\cite{Zhang25}; however, this is different from the CVT result, for which the magnitudes of $\rho_a$ and $\rho_b$ are flipped in this hierarchy.

\begin{table}
    \begin{center}
    \caption{Polynomial fit parameters used to model bar geometry resistivity data $\rho(T)$ for MSF-grown \ute samples.}
    \begin{tabular}{ | c | c | c | c | }
     & $\rho_a$ & $\rho_b$ & $\rho_c$ \\ 
    \hline
    $A\ (\mu\Omega\ m)$ & $8.02\EE{-2}$ & $8.89\EE{-2}$ & $1.08$ \\
    $B\ (\mu\Omega\ m/K^2)$ & $1.28\EE{-2}$ & $1.27\EE{-2}$ & $0.151$ \\
    $C_0\ (\mu\Omega\ m)$ & $-9.01\EE{-2}$ & $3.36$ & $34.3$ \\
    $C_1\ (\mu\Omega\ m/K)$ & $0.207$ & $-1.46$ & $-16.7$ \\
    $C_2\ (\mu\Omega\ m/K^2)$ & $-6.15\EE{-2}$ & $0.243$ & $3.24$ \\
    $C_3\ (\mu\Omega\ m/K^3)$ & $1.02\EE{-2}$ & $-1.51\EE{-2}$ & $-0.259$ \\
    $C_4\ (\mu\Omega\ m/K^4)$ & $-4.77\EE{-4}$ & $4.19\EE{-4}$ & $9.36\EE{-3}$ \\
    $C_5\ (\mu\Omega\ m/K^5)$ & $7.31\EE{-6}$ & $-4.27\EE{-6}$ & $-1.28\EE{-4}$
    \label{tab:UTe2_MSF_rho_poly}
    \end{tabular}
    \end{center}
\end{table}

\section{Other notes about high-loss modes}
\label{sec:high_loss}
In this appendix, we will make some additional remarks about the modes which make up the arc of high residual loss. We discuss the correspondence of the modes demonstrating an `exceptional' enhancement of $\sigma_1(T)$ at low-temperature with this region of induced-current weight parameter space (Appendix~\ref{sec:sigma}). We will also consider the possibility of two-level system (TLS) loss as a contribution to the large residual loss we observe (Appendix~\ref{sec:TLS}).

\begin{figure}[t!]	
\begin{subfigure}[b]{0.475\textwidth}
\caption{}
\includegraphics[scale=0.23,width=1.0\textwidth]{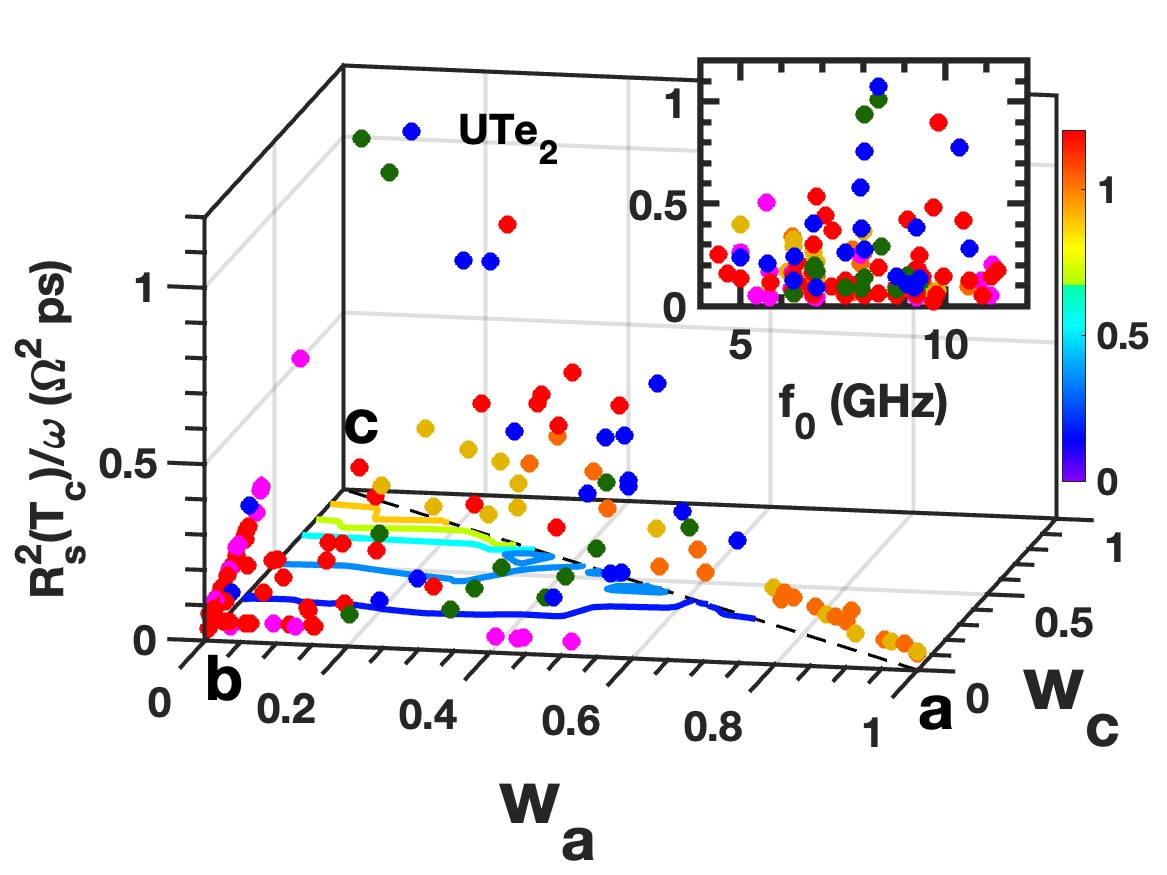}
\label{fig:Rs2_Tc_vs_w}
\end{subfigure}
\begin{subfigure}[b]{0.475\textwidth}
\caption{}
\includegraphics[scale=0.23,width=1.0\textwidth]{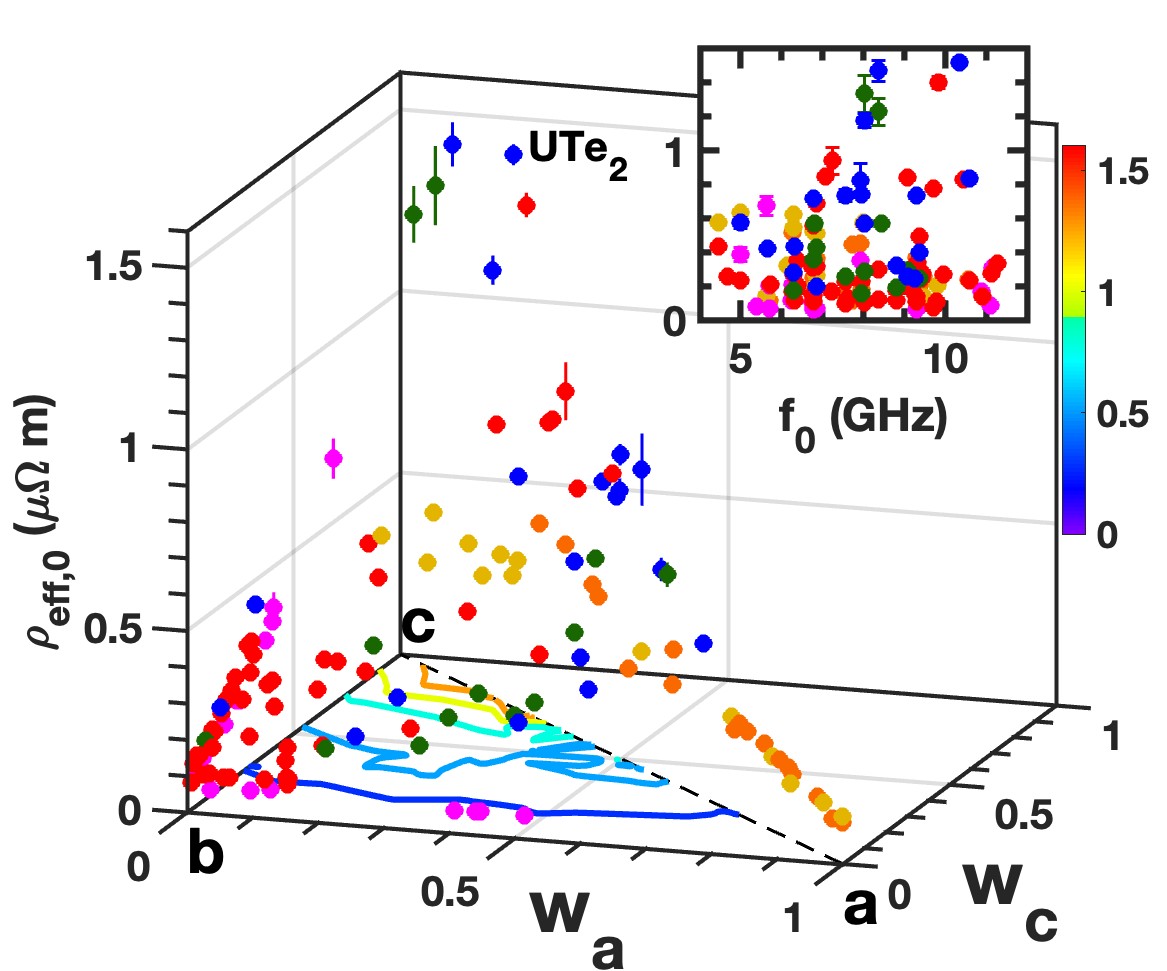}
\label{fig:rho_eff_0_vs_w}
\end{subfigure}
\caption{Plots vs the induced-current weightings (and vs mode frequency in the insets) of (a) a nominally frequency independent measure of the normal state residual loss (squared) and (b) the zero-temperature (in the absence of superconductivity) effective resistivity ($\rho_{\text{eff}} = \frac{2 R_s X_s}{\mu_0\omega}$) implied by our surface impedance data determined from normal state power-law fits.
Data for two CVT-grown \ute samples B39 and B40 is shown in orange and yellow, respectively. For the MSF-grown \ute samples, data for sample 23B is shown in magenta while data for sample 28 is shown in red, green, and blue for the cases of the $a$-, $b$-, and $c$-axes aligned with axis of the `hot finger', respectively. Contours of the data are also included in the $w_a$-$w_c$ plane with colors (indicated in the colorbars) corresponding to the value of the quantity being plotted.}
\label{fig:NS_loss_w_dep}
\end{figure}	

Firstly, we note that this arc of large residual loss is a superconducting state phenomenon. The normal state loss, shown in Fig.~\ref{fig:NS_loss_w_dep}, has a distinctly different dependence on the induced current weights. In the normal state, the loss is expected to scale as $\sim \sqrt{\omega}$, so we plot $R_s^2(T_c)/\omega$ in Fig.~\ref{fig:Rs2_Tc_vs_w} as a nominally frequency independent measure of the (square of the) normal state loss. This shows a clear planar increase from low loss for modes with currents in the $a$- and $b$-directions to high loss with currents in the $c$-direction. The character of this anisotropy arises from the anisotropy of the resistivity tensor. We determine a composite, effective resistivity $\rho_{\text{eff}} = \frac{2 R_s X_s}{\mu_0\omega}$ from our surface impedance data, which we fit to a power-law form given, 
\begin{equation}
    \frac{\rho_{\text{eff}}(T)-\rho_{\text{eff},0}}{\rho_{\text{eff,0}}} = \eta_{\rho_{\text{eff}}} \left(\frac{T}{T_c}\right)^{\alpha_{\rho_{\text{eff}}}},
    \label{eq:rho_eff_pow_law}
\end{equation}
in the normal state. We plot the zero-temperature value $\rho_{\text{eff},0}$ from this fit vs the induced-current weights in Fig.~\ref{fig:rho_eff_0_vs_w}, which demonstrates the same trend observed for $R_s^2(T_c)/\omega$. The values in the corners of the triangular plane formed by the data this plot imply the (axis-resolved) tensor components of $\rho_{\text{eff},0}$. These are roughly consistent with the zero-temperature values from quadratic fits to the MSF dc resistivity tensor data discussed in Appendix~\ref{sec:rho_data}, which are $\rho_{0,i} = (0.08,\ 0.09,\ 1)\ \mu$\ohm m ($A$ in Table~\ref{tab:UTe2_MSF_rho_poly}). While there are some modes in these plots along the $w_b=0$ line which depart from the overall trend, these are isolated, unlike the clear arc of high residual loss shown in Fig.~\ref{fig:Rs_0_vs_w}. The frequency dependencies of the quantities in Fig.~\ref{fig:NS_loss_w_dep} are shown in the insets. There is an apparent `tent arch' shape to these frequency dependencies (though is less clear for $\rho_{\text{eff},0}$) which peaks at $7$ GHz; however, only a small portion of data (comprising the upper limit) forms this frequency dependence, as opposed to the planar dependence on the induced-current weights which involves most of the modes.

\subsection{Complex conductivity and exceptional modes}
\label{sec:sigma}
We will now discuss our results for the complex conductivity $\sigma$, specifically the real part $\sigma_1$, which is a measure of the loss; however, this requires more analysis and assumptions to determine than the surface resistance $R_s$, which relies almost directly on the data. For most modes, we observe a monotonic decrease in $\sigma_1(T)$ as the temperature drops through the superconducting state, as shown in the inset of Fig.~\ref{fig:Zs_vs_T}, or a peak slightly below $T_c$ which gives way to an eventual decrease at low enough temperature. For a few `exceptional' modes, though, we observe a monotonic increase of $\sigma_1(T)$ with decreasing temperature. This results in comparable values of $\sigma_1$ and $\sigma_2$ in the superconducting state for these modes, which is unconventional. This exceptional temperature dependence was also observed for a CVT sample \cite{Seokjin21}.

\begin{figure}[]	
\begin{subfigure}[b]{0.475\textwidth}
\caption{}
\includegraphics[scale=0.23,width=1.0\textwidth]{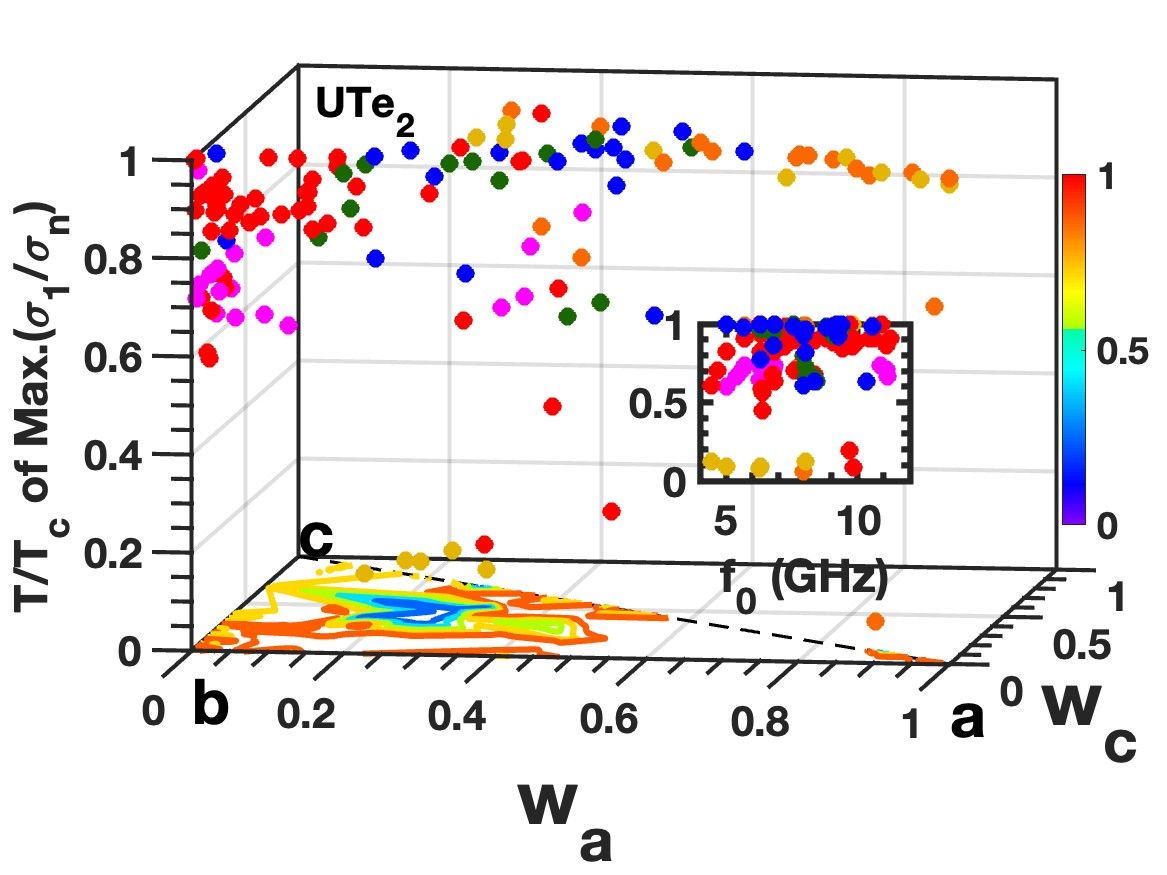}
\label{fig:t_sigma_1_peak_vs_w_sup_mat}
\end{subfigure}
\begin{subfigure}[b]{0.475\textwidth}
\caption{}
\includegraphics[scale=0.23,width=1.0\textwidth]{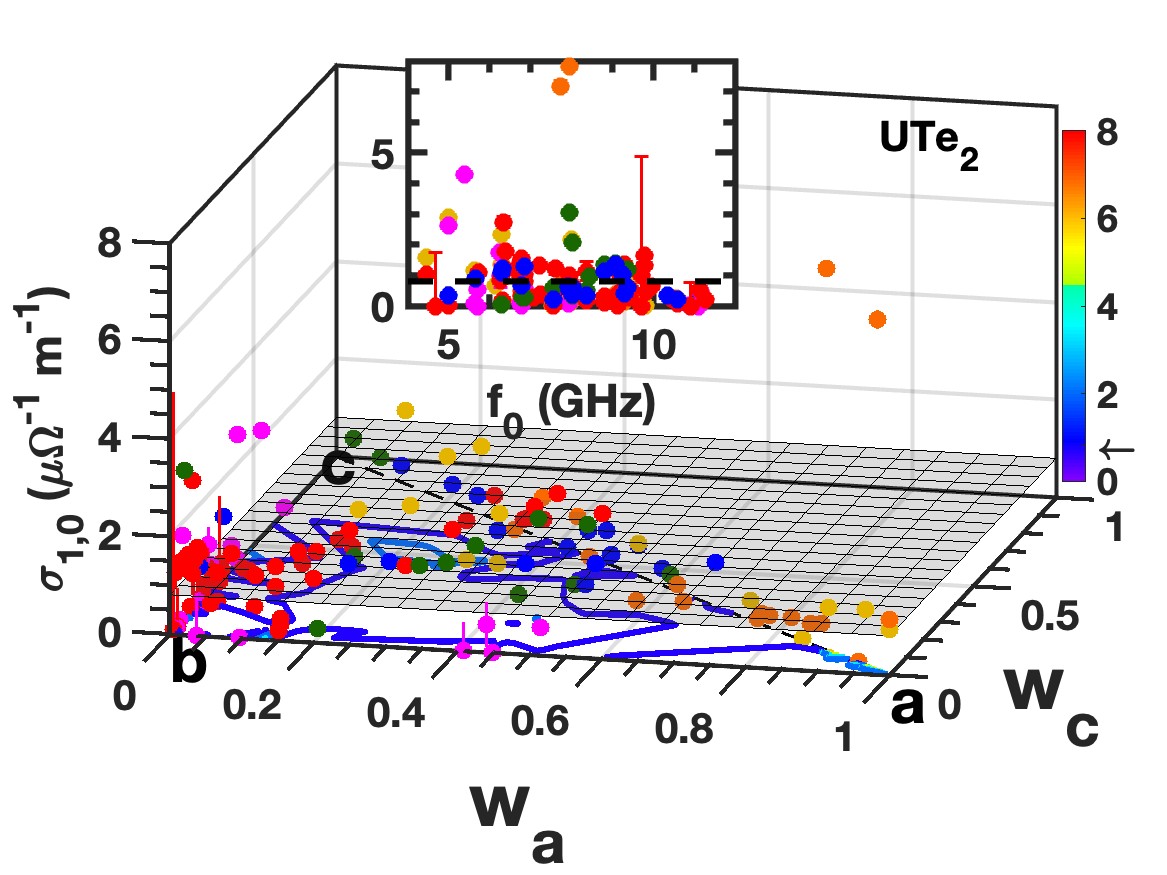}
\label{fig:sigma_1_0_vs_w_sup_mat}
\end{subfigure}
\caption{Plots vs the induced-current weightings (and vs mode frequency in the insets) of (a) the temperature (normalized by $T_c$) at which $\sigma_1$ attains its maximum below $T_c$ and (b) the residual value of $\sigma_1$ at zero-temperature.
Data for two CVT grown \ute samples B39 and B40 is shown in orange and yellow, respectively. For the MSF grown \ute samples, data for sample 23B is shown in magenta while data for sample 28 is shown in red, green, and blue for the cases of the $a$-, $b$-, and $c$-axes aligned with axis of the `hot finger', respectively. Contours of the data are also included in the $w_a$-$w_c$ plane with colors (indicated in the colorbars) corresponding to the value of the quantity being plotted.
Our estimate of the universal $d$-wave residual loss $\sigma_{00}$ is included in (b). This value is also marked with an arrow on the colorbar for the contours.}
\label{fig:sigma_peak_w_dep_sup_mat}
\end{figure}

To quantify the `exceptionality' of each mode, we plot the fraction of $T_c$ at which the maximum of $\sigma_1(T)$ occurs vs the induced-current weights in Fig.~\ref{fig:t_sigma_1_peak_vs_w_sup_mat}. Typical modes have values near $1$ in these plots while exceptional modes have values approaching $0$. A region of exceptional modes emerges in the induced-current weight space which can be seen as a blue region of the contours in Fig.~\ref{fig:t_sigma_1_peak_vs_w_sup_mat}. A top-down view of these contours is also shown in the inset of Fig.~\ref{fig:Rs_0_vs_w}, which demonstrates the correspondence between these exceptional modes and the arc of large residual loss, illustrated by a large, gray arrow in the main figure. The residual loss can also be exhibited through $\sigma_{1,0}$, the extrapolated zero-temperature value of $\sigma_1(T)$, as shown in Fig.~\ref{fig:sigma_1_0_vs_w_sup_mat}, for which this same region can be seen in the space of induced-current weight albeit less clearly than in the plot of $R_0$ in Fig.~\ref{fig:Rs_0_vs_w}. As discussed for $R_0$, we observe residual values of $\sigma_1(T)$ both above and below the universal $d$-wave loss $\sigma_{00}$, indicating that the scenario of line nodes is unlikely. The insets of Fig.~\ref{fig:sigma_peak_w_dep_sup_mat} show no particular frequency dependence of either the temperature of maximum $\sigma_1(T)$ below $T_c$ or the residual value of $\sigma_1(T)$.

We also note that there is a region near the $a$-direction with larger values of the frequency-independent loss ($R_0/\omega^2$ or $\sigma_{1,0}$) than those of the surrounding modes and for which the maximum of $\sigma_1(T)$ occurs at a lower temperature; however, this region is much smaller (only two points) than the arc and also has much smaller $R_0/\omega^2$ values than those of the arc. On the other hand, for $\sigma_{1,0}$, these two points have larger values than those of the rest of the arc; however, the arc of large residual loss is much less prominent in $\sigma_{1,0}$ to begin with.

\subsection{Input power and TLS loss}
\label{sec:TLS}

\begin{figure*}[ht!]  
\begin{subfigure}[b]{0.475\textwidth}
\caption{}
\hspace*{-0cm}\includegraphics[scale=0.6,width=1.0\textwidth]{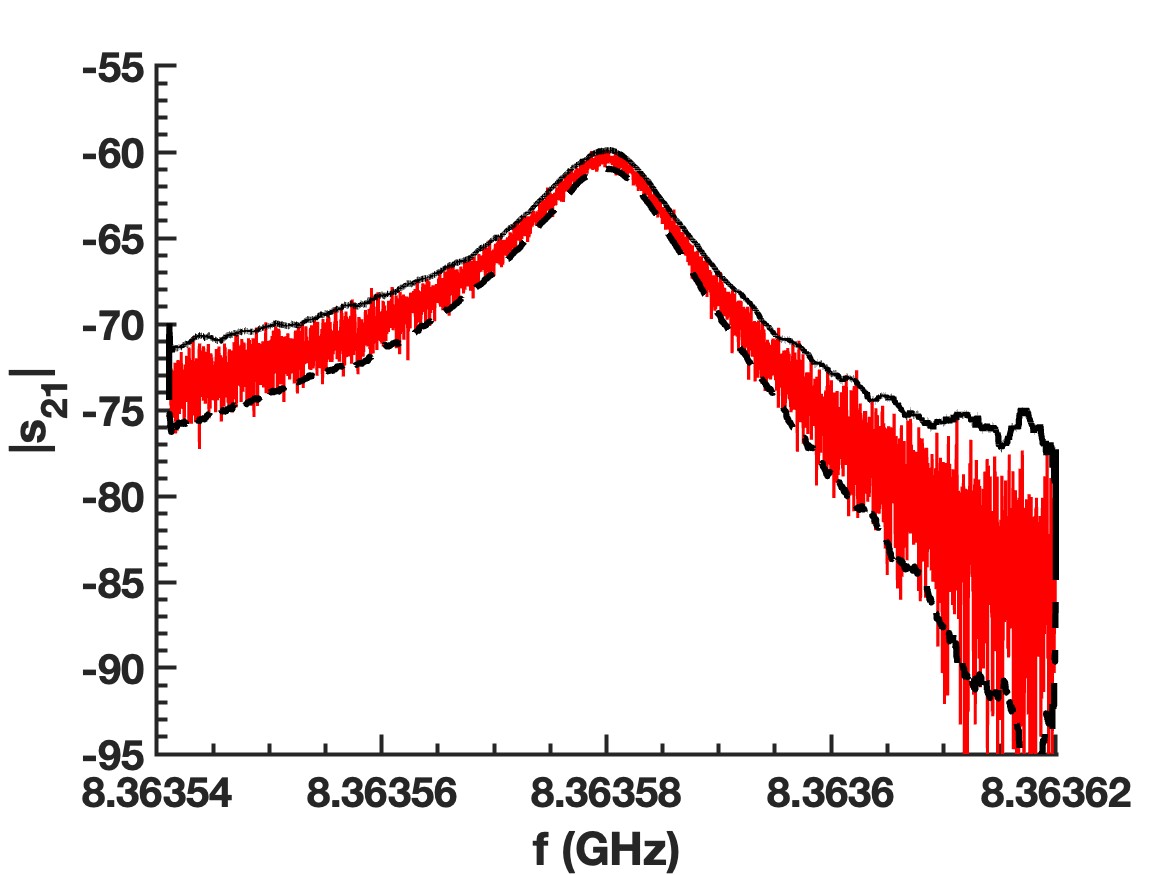}
\label{fig:bounding_alg}
\end{subfigure}
\begin{subfigure}[b]{0.475\textwidth}
\caption{}
\hspace*{-0cm}\includegraphics[scale=0.6,width=1.0\textwidth]{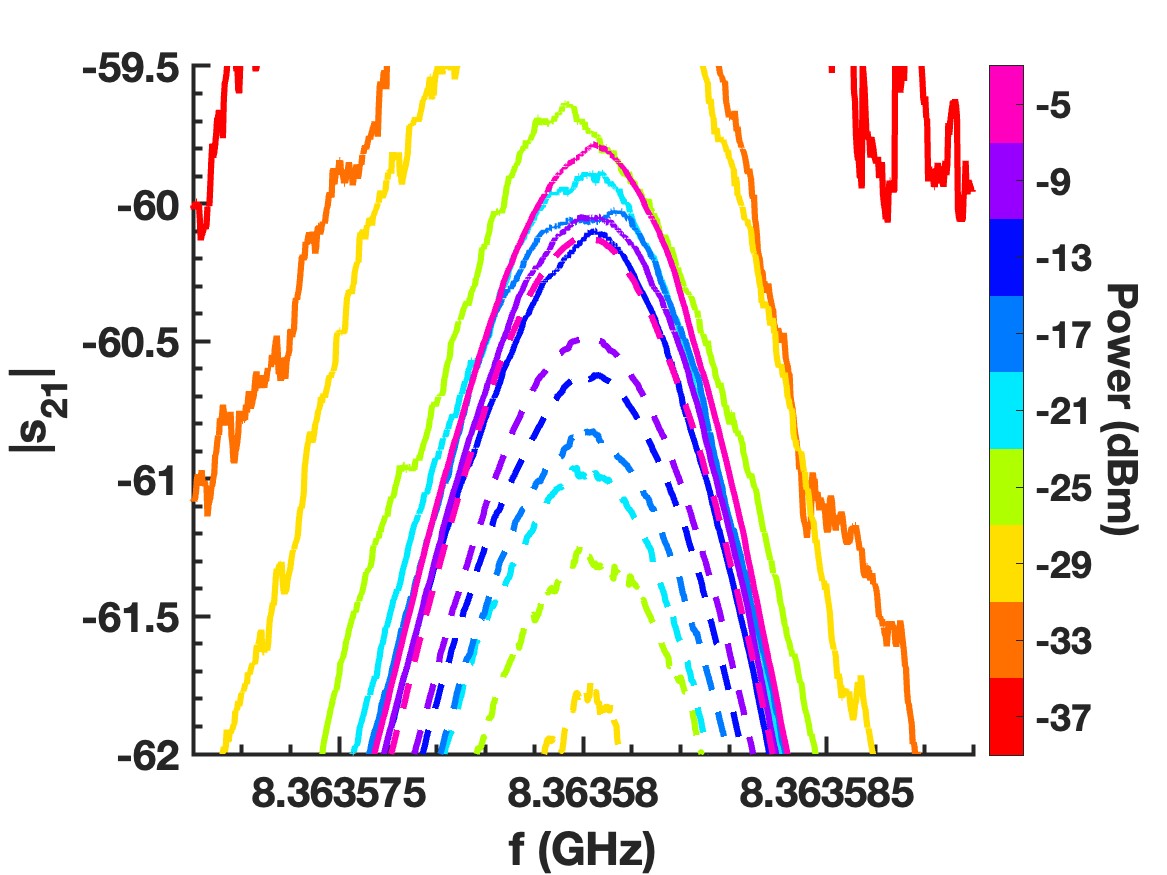}
\label{fig:shrinking_bounds_ex}
\end{subfigure}
\begin{subfigure}[b]{0.475\textwidth}
\caption{}
\hspace*{-0cm}\includegraphics[scale=0.6,width=1.0\textwidth]{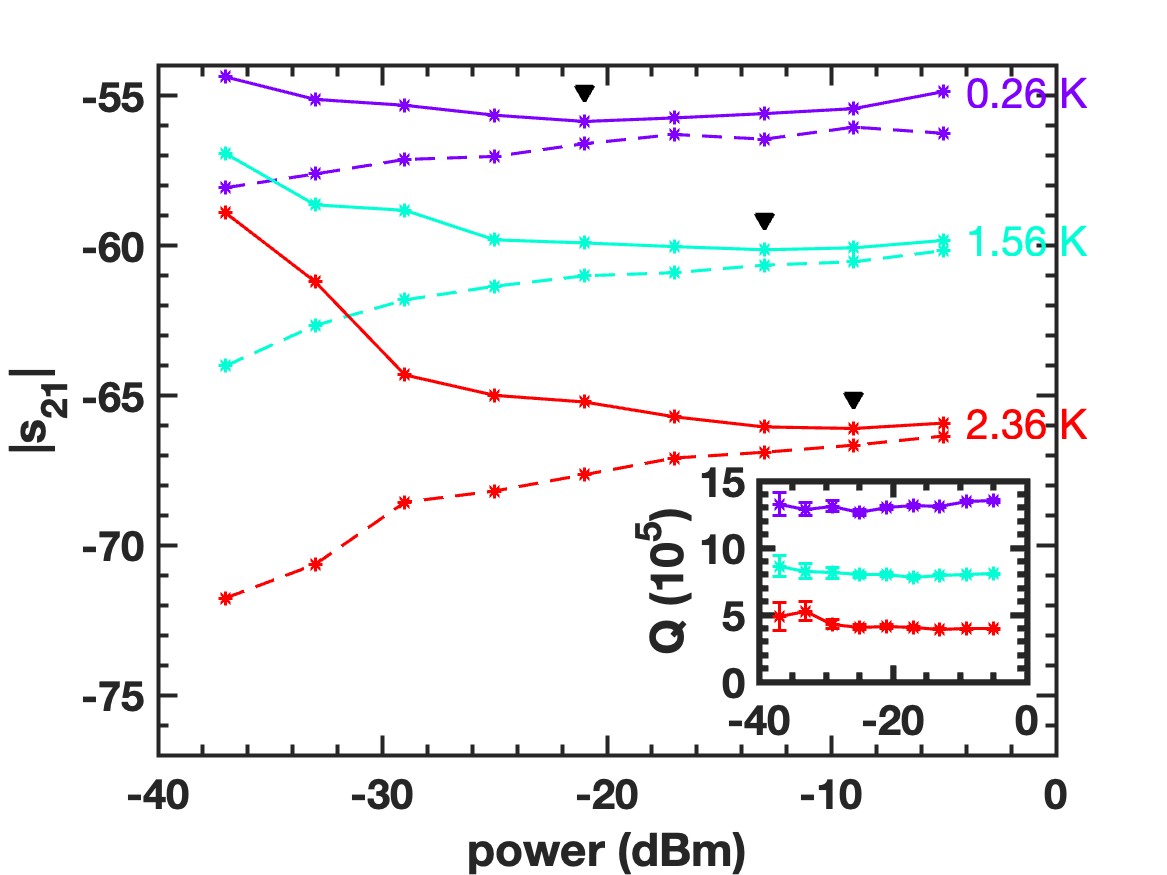}
\label{fig:bounds_pow_dep}
\end{subfigure}
\begin{subfigure}[b]{0.475\textwidth}
\caption{}
\hspace*{-0cm}\includegraphics[scale=0.6,width=1.0\textwidth]{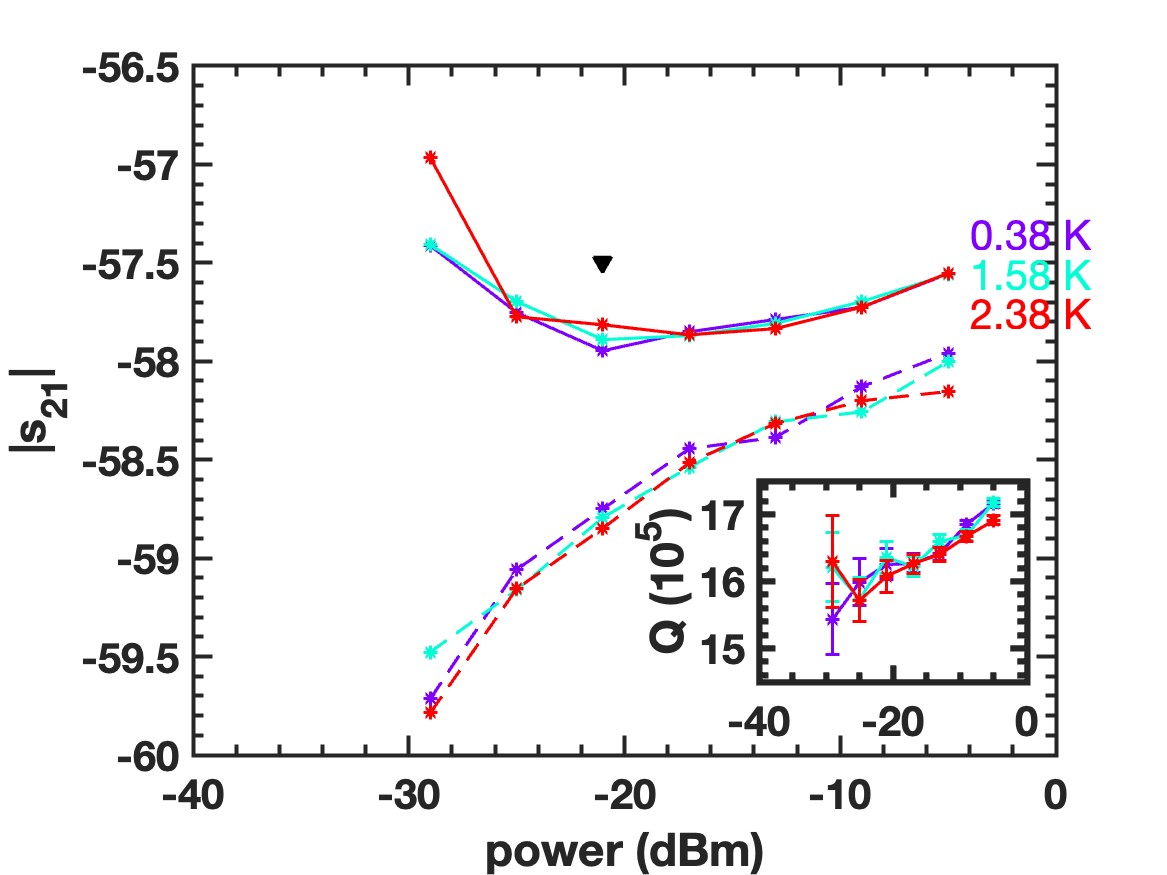}
\label{fig:bounds_pow_dep_BG}
\end{subfigure}
\caption{(a) An example of the raw transmission data $|S_{21}|$ in red and estimates of the upper (solid) and lower (dashed) bounds of the data in black. (b) Evolution of the bounds (solid for upper and dashed for lower) on $|S_{21}|$ near resonance with power increasing from red to pink. (c) Bounds of $|S_{21}|$ on resonance plotted vs power for three different temperatures. Optimal powers are marked for each temperature. (d) As in (c) but in the absence of a sample. The insets of (c) and (d) show the power dependencies of the corresponding quality factors for each temperatures.}
\label{fig:power_opt}
\end{figure*}	

The analyses performed in these experiments assume a linear response of the sample and the resonator components to the input rf fields. This means that a ratioed parameter such as $S_{21}$ does not depend on the input power, and thus, quantities derived from it, such as the surface impedance, are similarly independent of input power. Nonlinearities can arise from the response of the sample (due to the intrinsic nonlinearity of superconductors) or external/ background sources, such as an amplifier in the measurement circuit. It is assumed that for some range of input powers (generally low powers, as most physical systems will be pushed to nonlinearity with sufficiently high applied power), the response will be linear. The input power must be carefully selected to achieve a linear response. We find that this power varies greatly between resonant modes and can also vary with temperature. We have inspected the power dependence of each mode we study from the perspective of ensuring a linear response; however, these power dependencies have other significance, such as in determining the influence of two-level system (TLS) loss.

At low (input) powers, the transmission is noisy and may not be sufficiently prominent above the noise floor, but as the power increases, the signal will emerge from the noise floor where the output power was previously too low. Additionally, a higher input power can clean up the transmission by reducing noise in the measurement of $S_{21}$. It is, thus, favorable to use as high of a power as possible to get an adequate signal to noise ratio; however, too high of an input power will result in nonlinearity. The optimal power is then the maximum for which the response is still linear. Other components, such as amplifiers, may impose additional restrictions on acceptable input powers to avoid damage to equipment. As the input power is raised, the noise near resonance will decrease but the (true) shape of the resonance will not change until the optimal power, above which the resonance will depart from its low power behavior. 

To understand at what power $|S_{21}|$ begins to change shape, we carefully observe how upper and lower extents of the $|S_{21}|$ vs frequency curve change with input power near resonance. We determine approximate curves that bound the noisy $|S_{21}|$ data as shown in Fig.~\ref{fig:bounding_alg}. We determine a local mean and standard deviation of the $|S_{21}|$ data at each frequency point. This local mean is just the smoothed $|S_{21}|$ trace. The upper and lower bounds are set at two standard deviations above and below the mean, respectively. The reduction of noise due to increasing power can then be be seen in Fig.~\ref{fig:shrinking_bounds_ex}. As the power increases from red to pink, the upper (solid) and lower (dashed) bounds move together while staying inside the bounds of the previous power, up until the dark blue curve (-13 dBm). Above this power, the upper bound begins to increase as seen in the purple and pink curves. Figure~\ref{fig:bounds_pow_dep} plots the bounds (upper-solid, lower-dashed) on the transmission on resonance vs power for multiple temperatures. In this plot, the minimum in the upper bound (which is marked with a black triangle) indicates the optimal input power, which increases with temperature. The cyan curves, marked $1.56$ K, correspond to the examples in Figs.~\ref{fig:bounding_alg},\ref{fig:shrinking_bounds_ex}. The cyan upper bound of Fig.~\ref{fig:bounds_pow_dep} shows a minimum at $-13$ dBm, which agrees with the results of Fig.~\ref{fig:shrinking_bounds_ex}. These transmission measurements swept over various powers must be repeated at multiple temperatures throughout the superconducting state and into the normal state to determine the dependence of the linear regime on temperature. The purple upper bound ($0.26$ K) shows a minimum at $-21$ dBm, and the red upper bound ($2.36$ K) shows a minimum at $-9$ dBm. 

We have broadly observed the increase of the optimal power with temperature in the superconducting state for all the \ute samples we have studied. In the normal state, though, the optimal power remains mostly constant in temperature. In the absence of a sample, however, we have observed that the optimal power is fairly temperature independent. In Fig.~\ref{fig:bounds_pow_dep_BG}, the upper bounds at different temperatures for the corresponding background run all show a minimum near $-21$ dBm (which is marked); whereas, over roughly the same temperature range, Fig.~\ref{fig:bounds_pow_dep} indicated a significant departure from this optimal power at different temperatures with a sample present.

Above the optimal input power, $|S_{21}|$ tends to increase (both upper and lower bounds together). This is consistent with an increase in the quality factor (decrease in loss); however, the magnitude of the transmission at a sharp resonance can still be shifted up or down by changes in broad neighboring modes without necessarily affecting the $Q$ of the sharp mode. The power dependence of the quality factor at various temperatures is included in the insets of Figs.~\ref{fig:bounds_pow_dep}, \ref{fig:bounds_pow_dep_BG}. Even though for the sample run, $|S_{21}|$ on resonance is higher than it is for the background run at low temperatures, it is clear from the insets that $Q$ is consistently higher for the background run than for the sample run. The quality factor does indeed tend to increase with power (reduced loss at high powers) above the optimal power, though. While such behavior is suggestive of the two-level system (TLS) loss mechanism (which can commonly be hosted in dielectrics such as the sapphire and rutile elements used in the resonator), the change in $Q$ due to sweeping power is much smaller than that induced by changing temperature through the superconducting transition. TLS loss is, furthermore, expected to decrease with temperature, where a crossover between high loss at low temperature and low loss at high temperatures is expected to occur at temperatures around $\hbar\omega/k_B\sim 0.4$ K. For the background case shown in Fig.~\ref{fig:bounds_pow_dep_BG}, we do not observe the expected increase in $Q$ with respect to temperature for the TLS loss mechanism, and in Fig.~\ref{fig:bounds_pow_dep}, the loss dramatically increases with temperature due to the superconducting transition. The temperature of the sapphire `hot finger' and sample (if present), which is specified in Figs.~\ref{fig:bounds_pow_dep}, \ref{fig:bounds_pow_dep_BG}, is higher than that of the other dielectrics in the resonator, which are nominally held at $\sim 100$ mK. Despite this, we would still expect to see some increase of $Q$ with temperature due to the heat which does leak in to the resonator body and attached dielectrics. It is unlikely, then, that a TLS mechanism is responsible for the increase in $Q$ with power which we observe in our data for this range of input powers; furthermore, this effect produces a far smaller change in $Q$ than does the superconducting transition, so superconductivity should be dominant over this effect. Since this effect is observed across all samples we have studied as well as in the absence of a sample, it is very unlikely that it is due to the sample, so the background subtraction to isolate the resonant properties of the sample \cite{Mao95,Seokjin21} should remove any trace of this effect. The sample and background runs are performed at different powers by design to maintain a consistent linear response. The observed effect is likely due to nonlinearities of the output amplifier.

\bibliography{UTe2_MSF_bib.bib}

\end{document}